\documentclass[aps,twocolumn,showpacs,byrevtex,prl,reprint, nofootinbib]{revtex4-1}
\usepackage{xspace}
\usepackage{epsfig}
\usepackage{epstopdf}
\usepackage{dcolumn}
\usepackage{amssymb}   
\usepackage{amsmath}
\usepackage{mathrsfs}
\usepackage{bm}
\usepackage{xcolor}
\usepackage[normalem]{ulem}
\usepackage{graphicx}
\usepackage{subfigure}
\usepackage{siunitx}
\usepackage{pstricks}
\usepackage{multirow}
\usepackage{pst-node}
\usepackage{rotating}
\usepackage{times}
\usepackage{soul}
\usepackage{ulem}
\usepackage{balance}
\usepackage{lipsum}
\usepackage[english]{babel}
\addto{\captionsenglish}{%

}
\usepackage[colorlinks,linkcolor=blue, citecolor=blue]{hyperref}
\usepackage{lineno}
\usepackage[T1]{fontenc}
\usepackage{lmodern}
\usepackage{bm}
\newcommand{\BESIIIorcid}[1]{\href{https://orcid.org/#1}{\hspace*{0.1em}\raisebox{-0.45ex}{\includegraphics[width=1em]{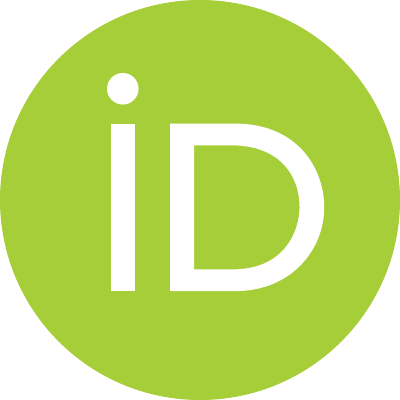}}}}
\newcommand{\EE}{e^+e^-}

\newcommand{\ar}{\rightarrow}
\newcommand{\llb}{\Lambda\bar{\Lambda}}

\newcommand{\KXSCC}{K^+\Xi^0\bar{\Sigma}^-}
\newcommand{\KXSCCbar}{K^-\bar{\Xi}^0\Sigma^+}
\newcommand{\KSXSCC}{K^0_{S}\bar\Xi^+\Sigma^{-}}
\newcommand{\KXSZCC}{K^-\bar\Xi^{+}\Sigma^{0}}

\RequirePackage[normalem]{ulem} 
\RequirePackage{color}\definecolor{RED}{rgb}{1,0,0}\definecolor{BLUE}{rgb}{0,0,1} 

\DeclareSIUnit{\bmm}{\bm{m}}
\DeclareSIUnit{\clight}{\textnormal{\textit{c}}}

\makeatletter
\def\ps@titlepage{%
  \let\@oddhead\@empty
}
\makeatother

\begin{document}
\normalsize
\parskip=5pt plus 1pt minus 1pt
\title{\protect\boldmath Measurement of Born cross sections for $e^+e^- \to \KXSCC$ at $\sqrt{s} = 3.51\text{--}4.95$~GeV and observation of $\psi(3770) \to \KXSCC$}
\author{BESIII Collaboration}
\thanks{Full author list given at the end of the Letter}

\begin{abstract}
Using 44.55~fb$^{-1}$ of $e^+e^-$ collision data collected by the BESIII detector at the BEPCII collider, we report the first measurement of the Born cross sections for the $e^+e^-\to\KXSCC$ reaction at 56 center-of-mass energies between 3.51 and 4.95~GeV.
A fit to the dressed cross sections reveals the first observation of the $\psi(3770) \to \KXSCC$ process, with a statistical significance of 6.0$\sigma$ including systematic uncertainties. 
This result represents the first observation of charmless three-body baryonic decay of a vector charmonium state above the open-charm threshold, providing important experimental input for studying the origin of the large non-$D\bar{D}$ component of the $\psi(3770)$ decay width.
No significant signals for other charmonium(like) states \textit{i.e.\@}, $\psi(4040)$, $\psi(4160)$, $Y(4230)$, $Y(4360)$, $\psi(4415)$, $Y(4500)$, $Y(4660)$ or $Y(4710)$ are observed, and the upper limits for the product of the branching fraction and the electronic partial width at the 90\% confidence level for each assumed charmonium(like) state are provided.
Additionally, the ratios of Born cross sections between this work and the previous measurements of $e^+e^-\to \KSXSCC$ and $\KXSZCC$ are provided, which can be used to validate theoretical predictions related to isospin symmetry.
\end{abstract}

\maketitle
The study of charmonium(like) states produced in $e^+e^-$ annihilation and decaying into baryonic final states is a key element to test quantum chromodynamics-related theory. The potential model~\cite{Barnes:2005pb} predicts six vector charmonium states in the energy region from 3.7 to 4.7~GeV, identified as the $1D$, $3S$, $2D$, $4S$, $3D$, and $5S$ states~\cite{Brambilla:2010cs}. In the past decades, an abundance of charmonium(like) vector states has been observed with masses above the open-charm threshold at $e^+e^-$ colliders. 
Three conventional charmonium states, \textit{i.e.\@}, $\psi(4040)$, $\psi(4160)$, and $\psi(4415)$~\cite{BES:2001ckj}, have been observed in open-charm final states. Five nonconventional charmoniumlike states, \textit{i.e.\@}, $Y(4230)$, $Y(4360)$, $Y(4500)$, $Y(4660)$, and $Y(4710)$, have been observed in hidden-charm final states via initial state radiation (ISR) processes at BaBar~\cite{BaBar:2005hhc, BaBar:2006ait, BaBar:2012hpr, BaBar:2012vyb} and Belle~\cite{Belle:2007dxy, Belle:2007umv, Belle:2008xmh, Belle:2014wyt, Belle:2013yex} experiments, or by direct production processes at CLEO-c~\cite{CLEO:2006ike} and BESIII~\cite{BESIII:2014rja, BESIII:2022joj, BESIII:2023cmv, BESIII:2023wqy,BESIII:2024ths}.
These nonconventional states cannot be classified as resonances consisting solely of a pure $c\bar{c}$ quark pair.
The overpopulation of structures and the discrepancies between potential model predictions and experimental measurements suggest that some of these structures may be candidates for exotic states~\cite{Chen:2016qju,Wang:2025dur}. To explain their nature, many hypotheses including hybrid states~\cite{Briceno:2015rlt}, multiquark states~\cite{Close:2005iz}, and molecular states~\cite{WangQ:2014cvms} have been proposed. However, no definitive conclusion has been drawn. 
As suggested in Refs.~\cite{Wang:2019mhs,Qian:2021neg}, the observation of baryonic decay modes may favor an interpretation of some charmoniumlike states as predominantly conventional charmonium. Still, the coexistence of such decays with nonconventional features reflects our limited understanding of the strong interaction in the nonperturbative regime, calling for further experimental input.

The $\psi(3770)$ is generally interpreted as a conventional $c\bar{c}$ state located above the open-charm threshold and is expected to decay predominantly to $D\bar{D}$, with a branching fraction of at least 99\%~\cite{Eichten:1978tg}, but gluonic or light-quark degrees of freedom may enhance its decays to light-quark final states~\cite{Close:2005iz}. The first non-$D\bar{D}$ decay, $\psi(3770)\to\pi^+\pi^-J/\psi$, was observed by BES in 2003~\cite{BES:2003bes}, later confirmed by CLEO-c, which also reported other non-$D\bar{D}$ modes~\cite{CLEO:2005zky,CLEO:2005vma,CLEO:2006nor} and the first observation of light-hadron mode $\psi(3770)\to\phi\eta$~\cite{CLEO:2005zrs}. 

The measurement of the exclusive Born cross sections for the $e^+e^-\to \KXSCC$ process at center-of-mass (c.m.)~energies above the open-charm threshold offers the opportunity to search for the baryonic charmless decays of the vector charmonium(like) states.
Although experimental studies in this energy region have been performed by the BESIII experiments~\cite{Ablikim:2013pgf,BESIII:2017kqg,Ablikim:2019kkp,  BESIII:2021ccp, BESIII:2021cvv,BESIII:2023rse, BESIII:2023rwv,BESIII:2024umc,BESIII:2024ogz,BESIII:2024gql}, only some evidence for $\psi(3770)\to\llb$, $\psi(3770)\to\Xi^-\bar\Xi^+$ and $\psi(4160) \to K^-\bar{\Xi}^+\Lambda$~\cite{BESIII:2021ccp, BESIII:2023rse, BESIII:2024ogz} processes have been reported. No significant signals of baryonic decays for other vector charmonium(like) states have been found.

In this paper, we present a measurement of the Born cross sections for the reaction $e^+e^- \to \KXSCC$, in the range of c.m.~energy ($\sqrt{s}$) from 3.51 to 4.95~GeV~\cite{BESIII:2015zbz, BESIII:2020eyu}.
Unless otherwise noted, the charge-conjugate process $e^+e^-\to\KXSCCbar$ is included by default.
This measurement is based on $e^+e^-$ collision data corresponding to a total integrated luminosity of 44.55~fb$^{-1}$~\cite{BESIII:2015qfd,BESIII:2022dxl,BESIII:2022ulv,BESIII:2024lbn,BESIII:2024lks} collected by BESIII detector~\cite{besiii} at the BEPCII collider~\cite{BEPCII}. 
Potential resonances, \textit{i.e.\@}, $\psi(3770)$, $\psi(4040)$, $\psi(4160)$, $Y(4230)$, $Y(4360)$, $\psi(4415)$, $Y(4500)$, $Y(4660)$ and $Y(4710)$ are studied by fitting the dressed cross sections of the $e^+e^- \to \KXSCC$ reaction.
The $\psi(3770) \to \KXSCC$ process is observed for the first time with a significance of 6.0$\sigma$ including systematic uncertainties.
In addition, the ratios of Born cross sections of the $e^+e^-\to\KXSCC$ reaction to those of the two isospin partner processes $e^+e^-\to \KSXSCC$ and $\KXSZCC$ are provided.

The BESIII geometric description and the detector response are modeled with a {\sc geant4}-based software package~\cite{GEANT4}.
The subsequent decays for $\Xi^0\to\Lambda\pi^0$ and $\pi^0\to\gamma\gamma$ are processed via the {\sc evtgen} generator~\cite{evtgen2,EVTGEN} according to the branching fractions provided by the Particle Data Group (PDG)~\cite{ParticleDataGroup:2022pth}.
The detection efficiency is determined by Monte Carlo (MC) simulations with a sample of $10^5$ events at each c.m.~energy point, with a phase-space model by the {\sc kkmc} generator~\cite{KKMC} including effects of the beam energy spread and ISR effect.
To estimate the background, an inclusive MC sample is generated at a c.m.~energy of 3.773~GeV. The inclusive MC sample includes the production of $D\bar{D}$ pairs (including quantum coherence for the neutral $D$ channels), the non-$D\bar{D}$ decays of the $\psi(3770)$, the ISR production of the $J/\psi$ and $\psi(2S)$ states, and the continuum processes incorporated in {\sc kkmc}generator~\cite{KKMC}.

Candidates for the process $e^+e^-\to\KXSCC$ are selected via a partial-reconstruction technique.
For each event, only the information of $K^{+}$ and $\Xi^0$ is measured, where $\Xi^0$ is reconstructed via the $\Lambda\pi^0$ mode with the subsequent decays $\Lambda\to p\pi^-$ and $\pi^0\to\gamma\gamma$, while the information of $\bar\Sigma^-$ is inferred from the recoiling system against the reconstructed $K^{+}\Xi^0$ system.

Charged tracks are reconstructed in the multilayer drift chamber (MDC) within the angular region $|\cos\theta| < 0.93$, where $\theta$ is 
the polar angle with respect to the $z$ axis in the laboratory system, which is the MDC symmetry axis.
At least two positive and one negative charged tracks are required to be reconstructed in the MDC.
Particle identification~(PID) for charged tracks combines measurements of the specific ionization energy loss in the MDC~(d$E$/d$x$) and the flight time in the time-of-flight (TOF) detector to form likelihoods $\mathcal{L}(h)~(h=p,K,\pi)$ for each hadron $h$ hypothesis.
Tracks are identified as protons when the proton hypothesis has the greatest likelihood ($\mathcal{L}(p)>
\mathcal{L}(K)$ and $\mathcal{L}(p)>\mathcal{L}(\pi)$), while charged pions are identified by requiring that $\mathcal{L}(\pi)>\mathcal{L}(p)$ and $\mathcal{L}(\pi)>\mathcal{L}(K)$, and the charged kaons are identified by requiring that $\mathcal{L}(K)>\mathcal{L}(\pi)$ and $\mathcal{L}(K)>\mathcal{L}(p)$.
Events with at least one proton, one positive kaon, and one negative pion are kept for further analysis.

Photons are reconstructed from isolated showers in the electromagnetic calorimeter (EMC). 
The energy deposited in the nearby time-of-flight counter is included to improve the reconstruction efficiency and energy resolution. 
The energies of photons are required to be greater than 25~MeV in the EMC barrel region ($|\cos\theta|<0.8$), and greater than 50~MeV in the EMC end cap ($0.86<|\cos\theta|<0.92$).
Furthermore, to suppress electronic noise and energy deposits unrelated to the collision events, the difference between the EMC time and the event start time is required to be within $0 < t < 700$~ns. 
To eliminate showers from charged tracks, the opening angle between the position of each shower in the EMC and any charged track must be greater than 10$^\circ$. 
Events with at least two photons are kept for further analysis.

The $\pi^0$ candidates are reconstructed via a one-constraint (1C) kinematic fit by looping over all $\gamma\gamma$ combinations by constraining $M_{\gamma\gamma}$ to $M_{\pi^0}^{\rm PDG}$, where $M_{\gamma\gamma}$ is the invariant mass of the $\gamma\gamma$ combination, and $M_{\pi^0}^{\rm PDG}$ is the nominal $\pi^0$ mass taken from the PDG~\cite{ParticleDataGroup:2022pth}.
All $\gamma\gamma$ combinations with a successful fit ($\chi_{\rm 1C}^2 < 500$) are used to reconstruct the $\pi^0$.
The $\Lambda$ candidates are reconstructed via a vertex fit and secondary vertex fit~\cite{Xu:2009zzg} by looping over all $p\pi^-$ combinations.
Each candidate with the best $\chi^2$ is selected to reconstruct the $\Xi^0$.
To suppress background events, a mass window condition $|M_{p\pi^-} - m_{\Lambda}|\leq5$~MeV/${c^2}$ is imposed by optimizing the figure of merit, defined as $\mathcal{S}/\sqrt{\mathcal{S} + \mathcal{B}}$. Here, $M_{p\pi^-}$ denotes the invariant mass of the $p\pi^-$ combination, $m_{\Lambda}$ is the nominal $\Lambda$ mass taken from the PDG~\cite{ParticleDataGroup:2022pth}, $\mathcal{S}$ is the number of surviving signal events from the signal MC sample, and $\mathcal{B}$ is the number of surviving background events from the inclusive MC sample.
To further suppress background events, a requirement on the decay length ratio $(L/\delta L) > 2$ is imposed, where $L$ is the measured decay length of the $\Lambda$ particle and $\delta L$ represents its corresponding uncertainty.   
The $\Xi^0$ candidate is reconstructed by selecting the $\Lambda \pi^0$ combination that minimizes $\Delta M=|M^{\rm PDG}_{\Xi^0}-M_{\Lambda\pi^0}|< 10$~MeV/$c^2$ over all $\Lambda\pi^0$ combinations, where $M_{\Lambda\pi^0}$ is the invariant mass of the $\Lambda\pi^0$ combination and $M^{\rm PDG}_{\Xi^0}$ is the nominal $\Xi^0$ mass taken from the PDG~\cite{ParticleDataGroup:2022pth}. 
The $\bar\Sigma^-$ candidates are inferred from the mass recoiling against the $K^+\Xi^0$ system:
\begin{equation}
    M_{K^+\Xi^0}^{\rm recoil} = \sqrt{(\sqrt{s}-E_{K^+\Xi^0})^2-\lvert \boldsymbol{p}_{K^+\Xi^0}\rvert^2},
\end{equation}
where $E_{K^+\Xi^0}$ and $\boldsymbol{p}_{K^+\Xi^0}$ are the energy and momentum of the selected $K^+\Xi^0$ system in the $e^+e^-$ c.m.~system, respectively.
After applying the event selection criteria, the remaining background candidates predominantly originate from processes with a final-state topology similar to that of the signal channel, such as  
$e^+e^- \to K^{*+} \bar{p} \Lambda$, $K^{+} \bar{\Sigma}^{0} \Xi^{-}$, and $K^{+} \bar{\Lambda} \Xi^{-}$.
However, these background channels are found to have a smooth distribution in the signal region of $M^{\rm{recoil}}_{K^+\Xi^0}$.
\begin{figure}[h]
    \centering
    \hspace*{0.15em}
    \includegraphics[width=0.48\textwidth]{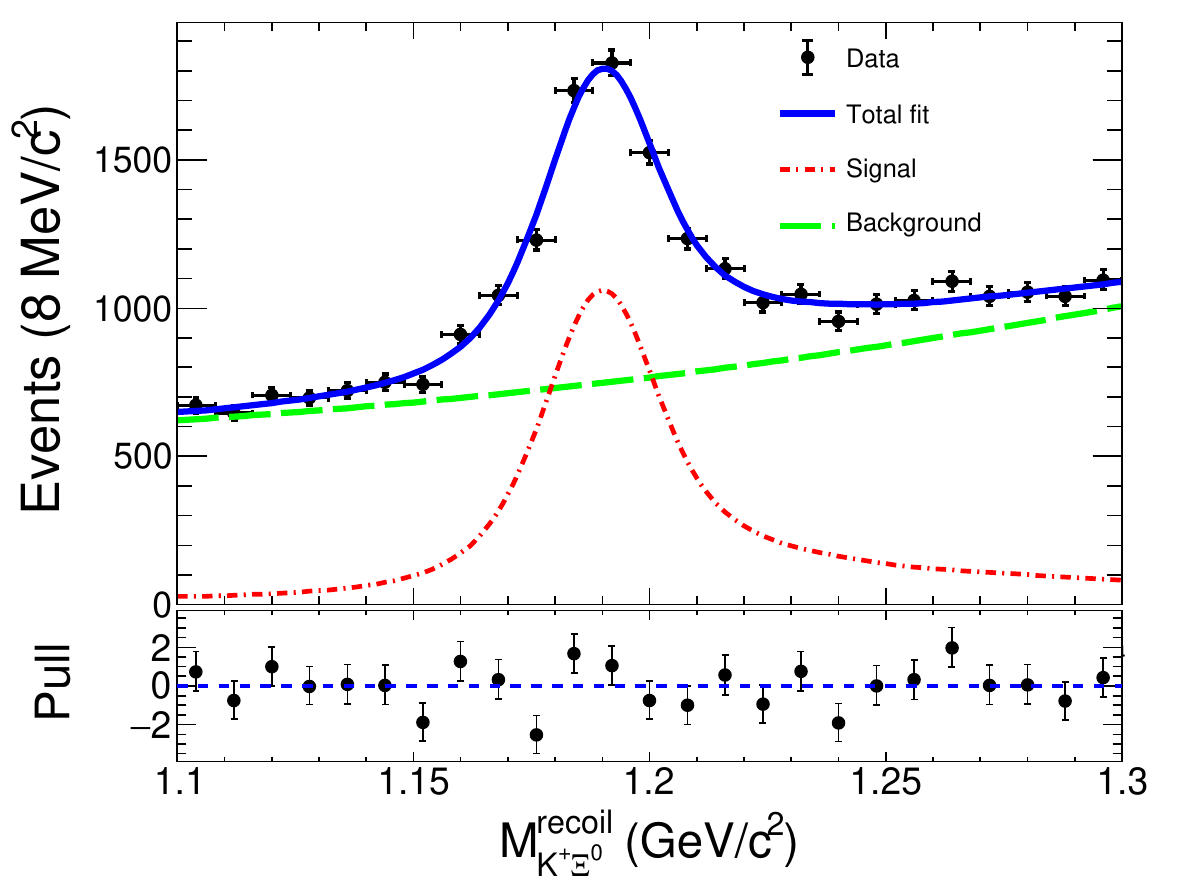}
    \caption{{Fit to $M^{\rm recoil}_{K^+\Xi^0}$ from the data sample at $\sqrt{s} = 3.773$~GeV. The dots with error bars are data, the blue solid line presents the fit result, the red dashed line is for signal shape, and the green dashed line stands for the background. The tail of signal is caused by ISR effect, which can be well simulated with KKMC.}}
    \label{datafit}
\end{figure}

The signal yield $N_{\rm obs}$ of the $\EE\to\KXSCC$ process at each c.m.~energy point is extracted by fitting the recoil mass spectrum of $K^+\Xi^0$ with an extended unbinned maximum likelihood in the range from 1.1 to 1.3~GeV/$c^2$ as shown in Fig.~\ref{datafit}.
In the fit, the signal shape is represented by the signal MC probability density function (PDF) convolved with a Gaussian function whose parameters are fixed using the combined sample of all energy points.
The background shape is described by a second-order Chebyshev polynomial function PDF.
The numerical results are summarized in the End Matter.
The Born cross section of the $\EE\to\KXSCC$ process at a given c.m.~energy is calculated by
\begin{equation}\label{BCS:ABC}
    \sigma^B = \frac{N_{\rm{obs}}}{
    \mathcal{L} \cdot (1+\delta) \cdot \frac{1}{\lvert 1-\Pi \rvert^2} \cdot \epsilon \cdot \mathcal{B}}~,
\end{equation}
where ${\cal{L}}$ is the integrated luminosity, $(1 + \delta)$ is the ISR correction factor, $\frac{1}{|1-\Pi|^2}$ is the vacuum polarization (VP) correction factor, $\epsilon$ is the detection efficiency, and ${\cal B}$ is the product of the branching fractions for $\Xi^0\to \Lambda\pi^0,~\pi^0\to \gamma\gamma$~and~$\Lambda \to p\pi^-$ taken from the PDG~\cite{ParticleDataGroup:2022pth}. The VP correction factor is calculated according to Ref.~\cite{Jegerlehner:2011ti}.
The initial value of the ISR correction factor is obtained using the QED calculation as described in Ref.~\cite{Kuraev:1985hb}. 
The detection efficiencies and ISR factors are determined iteratively following the methodology proposed in Ref.~\cite{Sun:2020ehv}.
The resulting Born cross sections for all c.m.~energy points are provided in Table~\ref{tab:signal:yields:DD}.
\begin{table*}
    \begin{center}
    \caption{
    The $e^+e^- \to \KXSCC$ born cross sections $\sigma^{B}$ for 56 energy points between 3.51 and 4.95~GeV. The values in the brackets are the corresponding upper limits at the 90\% confidence level. The first uncertainty is statistical, and the second systematic. The $\sqrt{s}$ is the c.m.~energy~\cite{BESIII:2015zbz, BESIII:2020eyu}. The $\int\mathcal{L}dt$ is the integrated luminosity~\cite{BESIII:2015qfd, BESIII:2022dxl, BESIII:2022ulv, BESIII:2024lbn, BESIII:2024lks}. The $N_{\rm S}$ is the observed signal yield. The VP correction factor $\frac{1}{|1-\Pi|^2}$, the $\epsilon (1+\delta)$ is the product of the ISR correction factor and the detection efficiency. The $Sig.(\sigma)$ is statistical significance.} 
    \begin{tabular}{l  l  l  l  l  l  l}
        \hline
        \hline
		$\sqrt{s}$ (GeV) &$\int \mathcal{L}$d$t \rm~(pb^{-1})$ &$N_{\rm S}$& $\frac{1}{|1-\Pi|^2}$ & $\epsilon (1+\delta)$  & $\sigma^{B}$ (fb)  &$Sig.(\sigma)$ \\
        \hline
3.510&404.7&$320^{+34}_{-33}$&1.045&0.152&$7879^{+839}_{-811}{\pm452}$&>10\\
3.514&40.9&$23.1^{+9}_{-23}~{(< 35.5)}$&1.044&0.147&$5821^{+2410}_{-2165}{\pm334}~{(< 8950)}$&2.9\\
3.553&42.173&$22.5^{+10}_{-23}~{(< 36.0)}$&1.041&0.144&$5648^{+2673}_{-2451}{\pm324}~{(< 9003)}$&2.5\\
3.554&129.4&$108^{+18}_{-17}$&1.041&0.140&$9067^{+1540}_{-1454}{\pm521}$&6.7\\
3.582&85.7&$47.9^{+13}_{-48}$&1.039&0.138&$6182^{+1769}_{-1641}{\pm355}$&4.3\\
3.650&410&$269^{+34}_{-34}$&1.021&0.140&$7278^{+930}_{-909}{\pm418}$&8.6\\
3.670&83.6&$10.4^{+10}_{-10}~{(< 23.5)}$&0.994&0.144&$1368^{+1440}_{-1368}{\pm79}~{(< 3104)}$&0.9\\
3.768&415.8&$78^{+20}_{-19}$&1.054&0.144&$1960^{+497}_{-471}{\pm113}$&4.1\\
3.773&20274.8&$6391^{+168}_{-168}$&1.059&0.136&$3481^{+92}_{-91}{\pm200}$&>10\\
3.780&410&$124^{+33}_{-26}$&1.057&0.145&$3131^{+835}_{-662}{\pm180}$&5.1\\
3.808&50.54&$40^{+10}_{-9}$&1.056&0.155&$7688^{+1946}_{-1782}{\pm441}$&4.9\\
3.867&108.9&$44^{+15}_{-16}$&1.051&0.153&$3982^{+1318}_{-1482}{\pm229}$&3.2\\
3.871&110.3&$81^{+20}_{-19}$&1.051&0.152&$7269^{+1817}_{-1718}{\pm417}$&3.6\\
3.896&52.61&$28^{+10}_{-10}$&1.049&0.148&$5532^{+1915}_{-1904}{\pm318}$&3.1\\
4.008&482&$166^{+29}_{-28}$&1.044&0.152&$3450^{+601}_{-584}{\pm197}$&6.0\\
4.086&52.86&$7.7^{+7}_{-8}~{(< 20.5)}$&1.051&0.150&$1507^{+1502}_{-1374}{\pm86}~{(< 3901)}$&1.1\\
4.128&401.5&$64^{+23}_{-20}$&1.052&0.134&$1801^{+653}_{-559}{\pm103}$&3.2\\
4.157&408.7&$79^{+21}_{-21}$&1.053&0.136&$2131^{+575}_{-555}{\pm121}$&3.6\\
4.178&3194.5&$631^{+73}_{-59}$&1.054&0.138&$2158^{+249}_{-203}{\pm123}$&>10\\
4.189&526.7&$134^{+24}_{-23}$&1.056&0.141&$2713^{+491}_{-474}{\pm155}$&5.4\\
4.199&526&$77^{+29}_{-22}$&1.056&0.142&$1554^{+590}_{-446}{\pm89}$&3.6\\
4.209&517.1&$108^{+23}_{-22}$&1.057&0.142&$2206^{+469}_{-450}{\pm126}$&5.6\\
4.226&514.6&$86^{+23}_{-30}$&1.057&0.148&$1702^{+457}_{-585}{\pm97}$&3.1\\
4.226&1100.9&$208^{+43}_{-32}$&1.056&0.146&$1946^{+407}_{-302}{\pm111}$&6.4\\
4.236&530.3&$122^{+24}_{-23}$&1.056&0.148&$2334^{+461}_{-440}{\pm133}$&4.7\\
4.242&55.88&$9.2^{+9}_{-9}~{(< 20.5)}$&1.055&0.143&$1725^{+1806}_{-1544}{\pm98}~{(< 3860)}$&1.1\\
4.244&538.1&$73^{+21}_{-23}$&1.056&0.141&$1446^{+424}_{-451}{\pm82}$&3.1\\
4.258&828.4&$158^{+36}_{-29}$&1.054&0.139&$2061^{+475}_{-382}{\pm117}$&5.5\\
4.267&531.1&$126^{+24}_{-28}$&1.053&0.145&$2458^{+465}_{-539}{\pm140}$&4.5\\
4.278&175.7&$43^{+16}_{-13}$&1.053&0.136&$2746^{+991}_{-826}{\pm156}$&3.5\\
4.288&502.4&$83^{+22}_{-28}$&1.053&0.133&$1875^{+506}_{-625}{\pm107}$&3.2\\
4.308&45.08&$10.2^{+7}_{-10}~{(< 23.0)}$&1.052&0.145&$2360^{+1706}_{-1578}{\pm134}~{(< 5315)}$&1.5\\
4.312&501.2&$69.5^{+22}_{-66}~{(< 99.5)}$&1.052&0.135&$1465^{+497}_{-478}{\pm83}~{(< 2222)}$&2.6\\
4.337&505&$112^{+24}_{-23}$&1.051&0.131&$2541^{+540}_{-520}{\pm145}$&3.8\\
4.358&543.9&$106^{+21}_{-23}$&1.051&0.142&$2074^{+404}_{-447}{\pm118}$&5.2\\
4.377&522.7&$60.3^{+22}_{-56}~{(< 117.5)}$&1.051&0.133&$1225^{+589}_{-472}{\pm70}~{(< 2553)}$&2.9\\
4.387&55.57&$1.0^{+13}_{-1}~{(< 15.5)}$&1.051&0.143&$177^{+1598}_{-177}{\pm10}~{(< 2943)}$&0.1\\
4.396&507.8&$54.9^{+20}_{-53}~{(< 97.0)}$&1.051&0.127&$1236^{+594}_{-497}{\pm70}~{(< 2268)}$&2.3\\
4.416&1090.7&$165^{+29}_{-28}$&1.052&0.138&$1647^{+287}_{-278}{\pm94}$&5.9\\
4.436&569.9&$81^{+23}_{-22}$&1.054&0.126&$1702^{+486}_{-457}{\pm97}$&3.6\\
4.467&111.09&$6.0^{+8}_{-7}~{(< 22.0)}$&1.055&0.135&$667^{+815}_{-667}{\pm38}~{(< 2203)}$&0.5\\
4.527&112.12&$24.3^{+11}_{-24}~{(< 47.0)}$&1.054&0.129&$2458^{+1307}_{-1123}{\pm140}~{(< 4894)}$&2.1\\
4.575&48.93&$12.6^{+8}_{-13}~{(< 19.5)}$&1.054&0.131&$2954^{+1903}_{-1738}{\pm168}~{(< 4578)}$&1.8\\
4.600&586.9&$65.7^{+20}_{-63}~{(< 93.0)}$&1.055&0.121&$1327^{+506}_{-431}{\pm75}~{(< 1969)}$&2.6\\
4.612&103.65&$8.1^{+7}_{-8}~{(< 22.0)}$&1.055&0.112&$1095^{+945}_{-916}{\pm62}~{(< 2862)}$&1.2\\
4.628&521.53&$66^{+18}_{-18}$&1.054&0.115&$1651^{+451}_{-461}{\pm94}$&3.1\\
4.641&551.65&$58^{+21}_{-19}$&1.054&0.116&$1368^{+484}_{-442}{\pm78}$&4.8\\
4.661&529.43&$40.0^{+19}_{-37}~{(< 64.5)}$&1.054&0.116&$913^{+556}_{-449}{\pm52}~{(< 1585)}$&1.7\\
4.682&1667.39&$105.2^{+33}_{-98}~{(< 138.0)}$&1.054&0.113&$780^{+370}_{-273}{\pm44}~{(< 1102)}$&2.6\\
4.699&535.54&$1.5^{+11}_{-18}~{(< 53.0)}$&1.055&0.115&$438^{+428}_{-438}{\pm25}~{(< 1291)}$&0.2\\
4.740&163.87&$15.3^{+10}_{-16}~{(< 29.0)}$&1.055&0.118&$1256^{+792}_{-733}{\pm71}~{(< 2262)}$&1.4\\
4.750&366.55&$4.7^{+13}_{-7}~{(< 32.5)}$&1.055&0.118&$230^{+446}_{-231}{\pm13}~{(< 1133)}$&1.3\\
4.781&511.47&$28.4^{+18}_{-26}~{(< 54.0)}$&1.055&0.123&$619^{+522}_{-426}{\pm35}~{(< 1290)}$&1.1\\
4.843&525.16&$41^{+18}_{-21}$&1.056&0.113&$1027^{+452}_{-519}{\pm58}$&4.2\\
4.918&207.82&$0.0^{+9}_{-0}~{(< 13.0)}$&1.056&0.111&$0^{+502}_{-0}{\pm0}~{(< 850)}$&0.0\\
4.951&159.28&$0.0^{+15}_{-0}~{(< 17.5)}$&1.056&0.111&$0^{+1030}_{-0}{\pm0}~{(< 1486)}$&0.0\\
        \hline
        \hline
    \end{tabular}
    \label{tab:signal:yields:DD}
    \end{center}
\end{table*}

The systematic uncertainties of the Born cross section measurement include contributions from the following sources: integrated luminosity, kaon tracking and PID, $\Xi^0$ reconstruction, fitting method, branching fractions of intermediate states, and the assumed line shape of the resonance fitting.  
The luminosity at each c.m.~energy point has been measured via Bhabha scattering events, with a systematic uncertainty of 1.0\%~\cite{BESIII:2015qfd} below 4.0~GeV, 0.7\%~\cite{BESIII:2022dxl} from 4.0 to 4.6~GeV, and 0.6\%~\cite{BESIII:2022ulv} above 4.6~GeV.
By using a control sample of $J/\psi \to K^{*}(892)^{0} K^{0}_{S}$ with the method from Ref.\cite{BESIII:2015dvj}, both the systematic uncertainties of kaon tracking and PID are estimated to be 1.0\%. 
The uncertainty due to the $\Xi^0$ reconstruction is evaluated to be 4.5\%
by incorporating the uncertainties of tracking and PID of the proton and pion, the $\Lambda/\Xi^0$ reconstruction, the decay length requirement of $\Lambda$, and the requirements of the $\Lambda$ and $\Xi^0$ mass windows using a control sample of $\psi(3686)\to\Xi^0\bar{\Xi}^0$~\cite{BESIII:2024ues}. 
The sources of the systematic uncertainty in the fit of the $M_{K^+\Xi^0}^{\rm{recoil}}$ spectrum cover the fitting range, signal shape, and background shape. 
The uncertainty due to the fitting range is evaluated by varying the nominal range of $M_{K_S^0\bar{\Xi}^+}^{\rm recoil}$ by 50~MeV/$c^2$, which contributes an uncertainty of 2.5\%.
The uncertainty of the signal shape for the $M_{K_S^0\bar{\Xi}^+}^{\rm recoil}$ fit is evaluated by combining all energy points and varying the Gaussian parameters within the range of 1$\sigma$. The resulting signal yield difference of 1.1\% is taken as the systematic uncertainty.
The uncertainty due to the background shape is estimated through an alternative fit with a lower-order polynomial function and a higher-order polynomial function PDF, and the larger value is taken as the result. The difference of 0.5\% from the nominal fit is taken as the systematic uncertainty.
The systematic uncertainty due to the branching fraction of the $\Lambda \to p\pi^-$ decay contributes $0.78\%$ as reported by the PDG~\cite{ParticleDataGroup:2022pth}, while the uncertainties from the branching fractions of $\Xi^0 \to \Lambda\pi^0$ ($<0.1\%$) and $\pi^0 \to \gamma\gamma$ ($<0.1\%$) decays are negligible.

The uncertainty of the input line shape includes two parts: the parametrization and the model of the line shape, which are evaluated using the method described in Ref.~\cite{BESIII:2024gql}. The first component arises from the statistical uncertainty of the input line shape of the cross sections. The uncertainty is estimated by varying the central values of the parameters according to their error matrix. Then, a Gaussian function is used to fit the $\epsilon\cdot(1+\delta)$ distribution. 
The second source stems from the influence of different resonances. For the $\psi(3770)$, since it is included in the nominal line shape, this uncertainty is estimated by varying the mass and width of the $\psi(3770)$ by 1$\sigma$ according to the PDG values~\cite{ParticleDataGroup:2022pth}. For other resonances, the uncertainty is quantified by incorporating the additional states in the ISR calculation.
The total systematic uncertainty is calculated to be 1.0\% by adding the three contributions in quadrature.

\begin{figure*}[!hbpt]
	\begin{center}
        \includegraphics[width=0.42\textwidth]{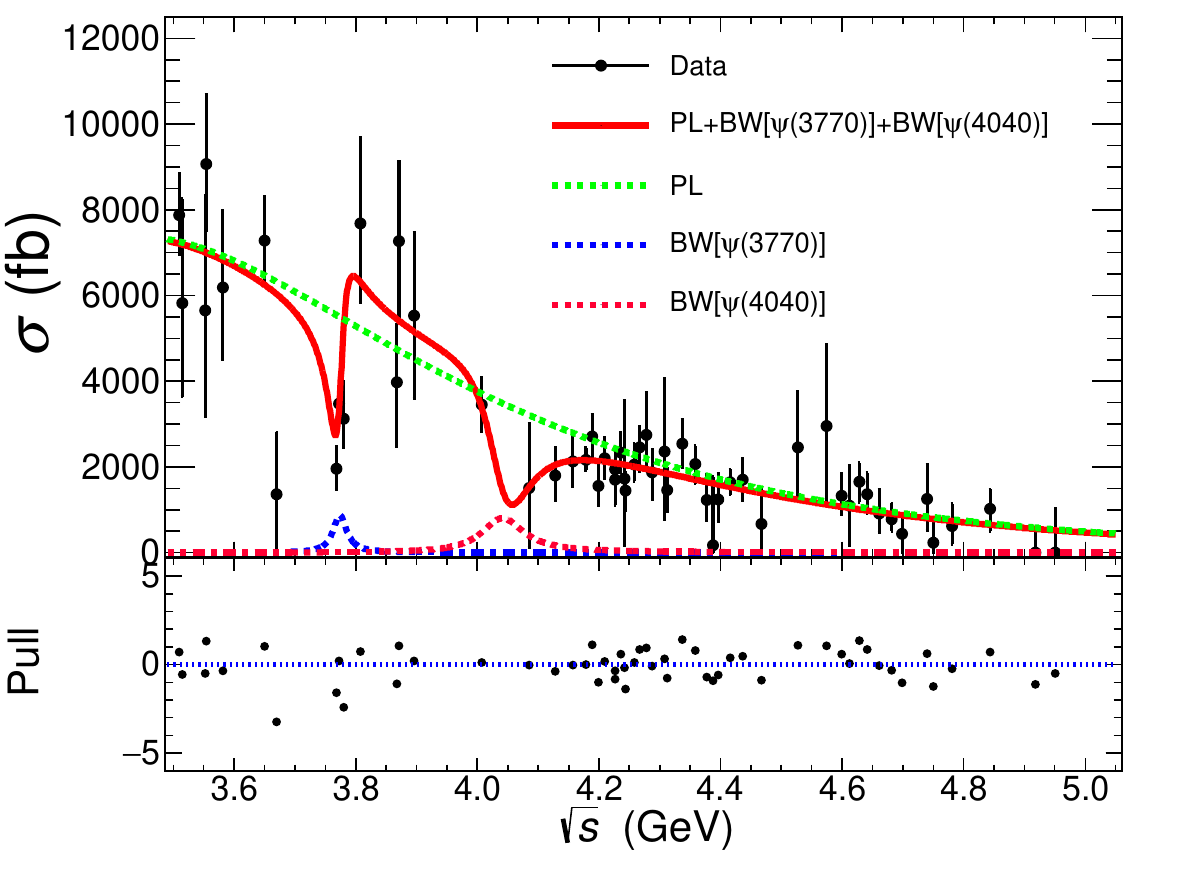}
        \includegraphics[width=0.42\textwidth]{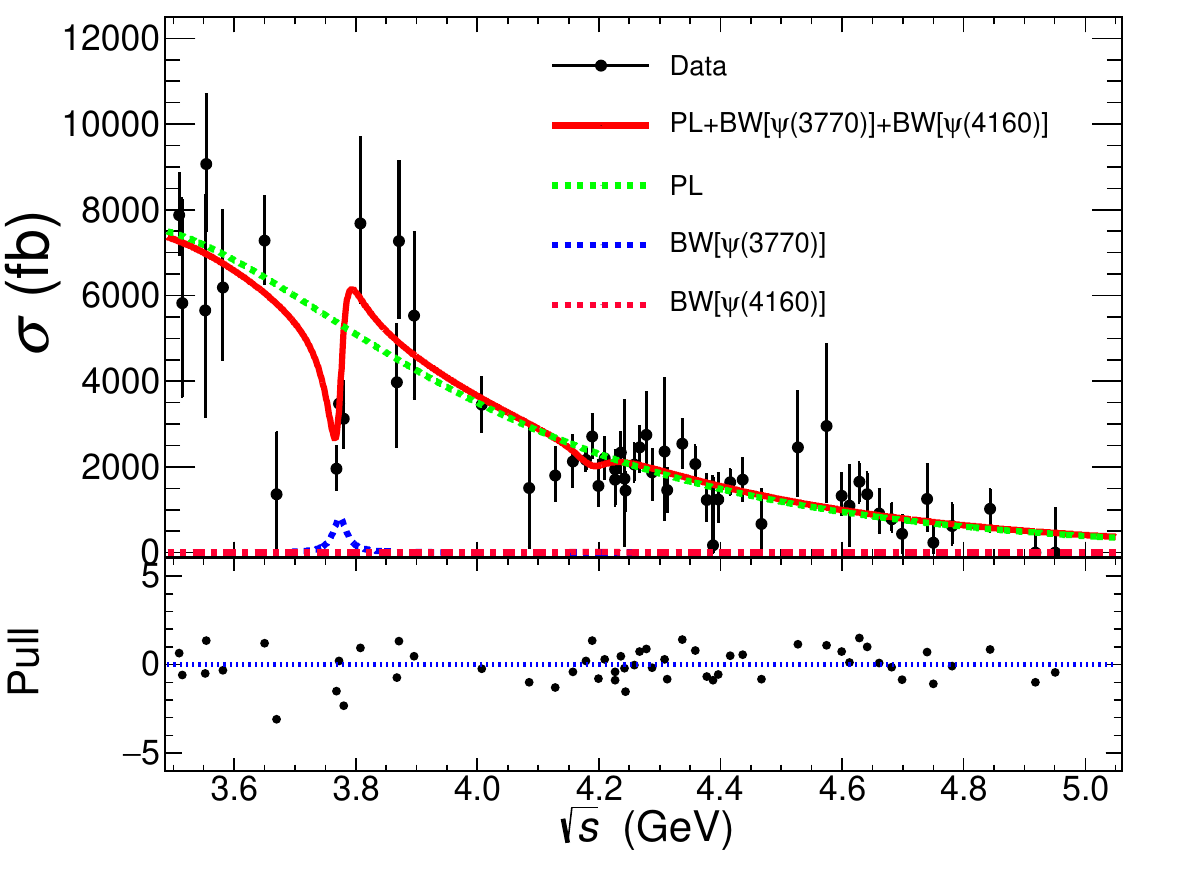}
        \includegraphics[width=0.42\textwidth]{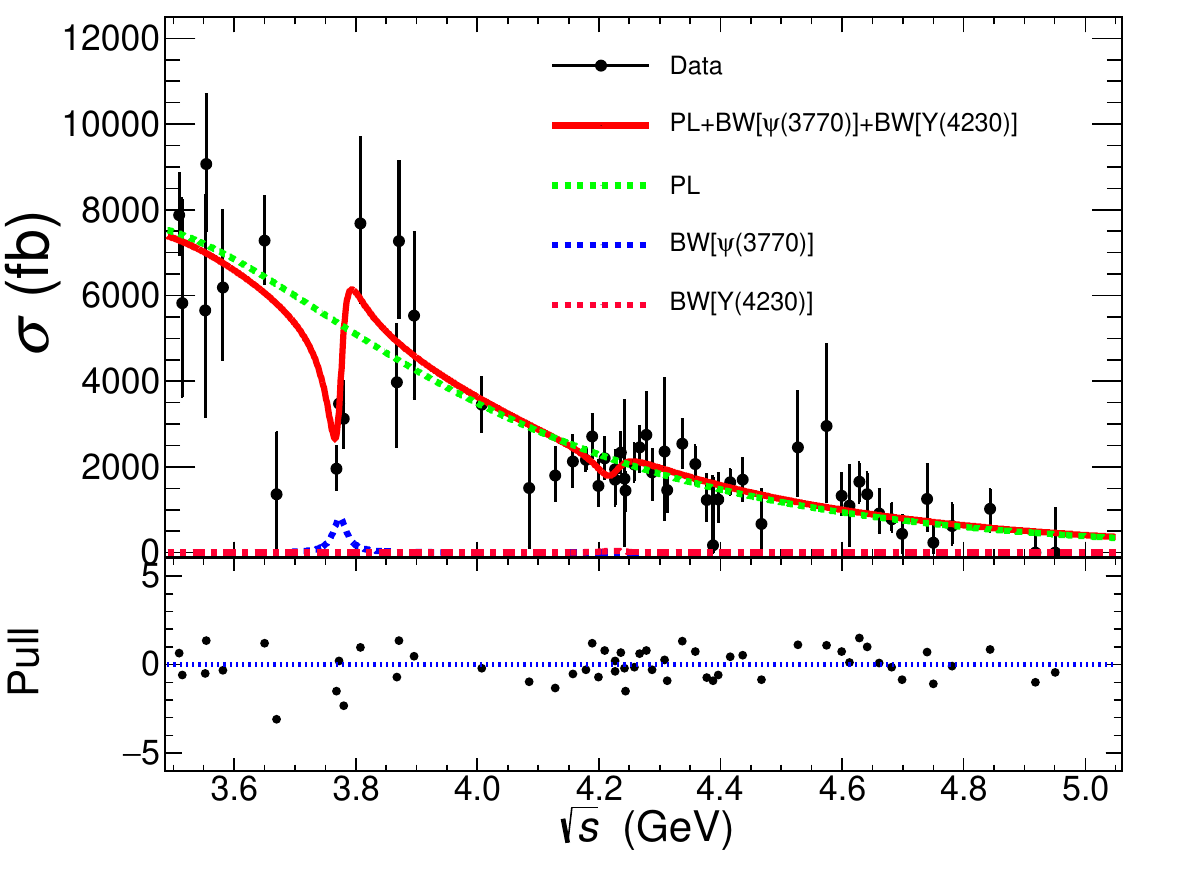}
        \includegraphics[width=0.42\textwidth]{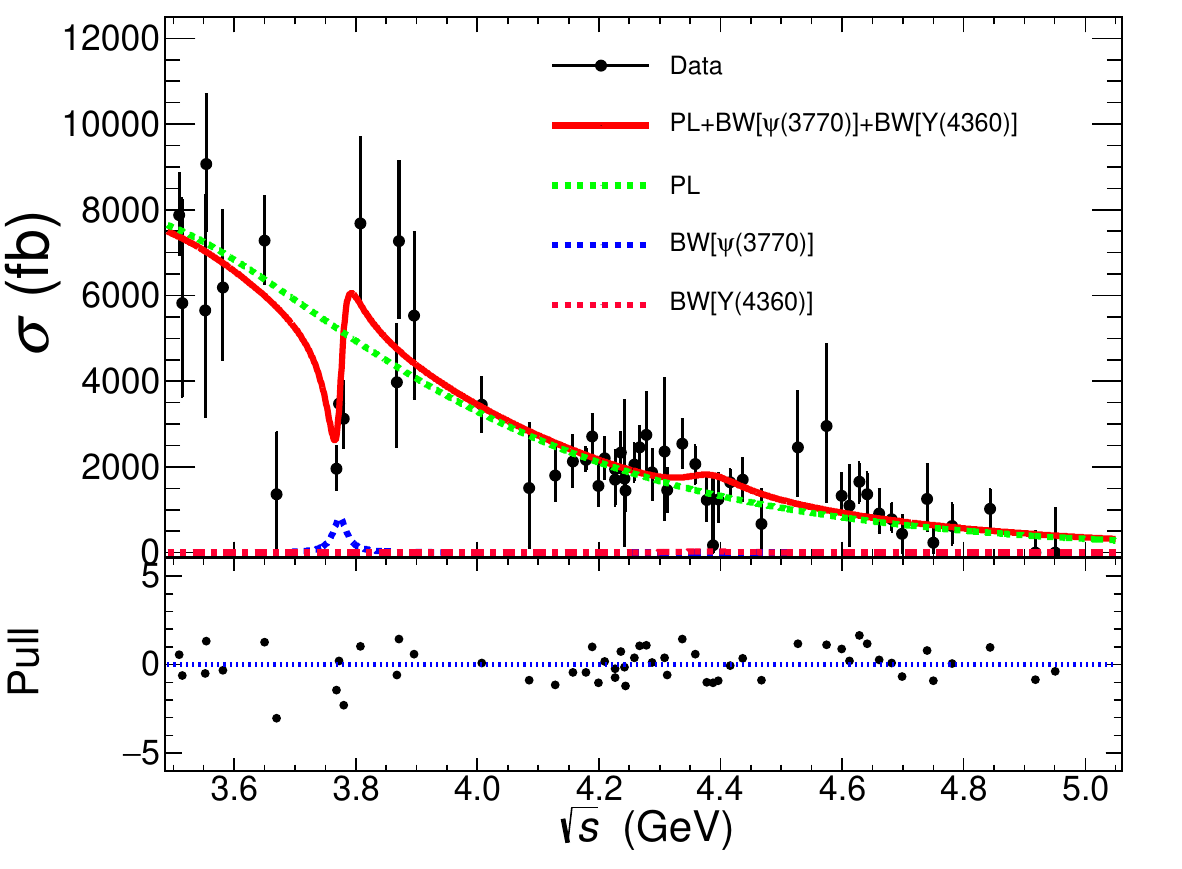}
        \includegraphics[width=0.42\textwidth]{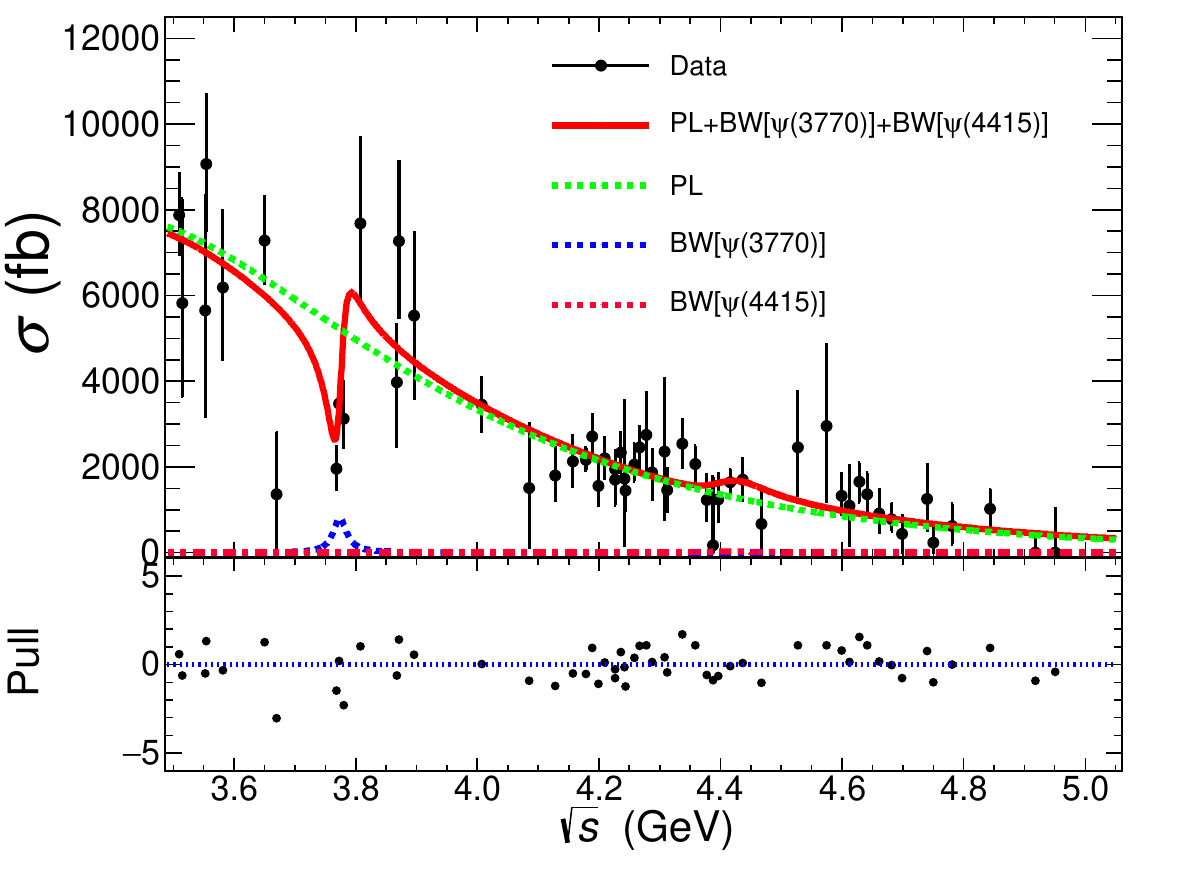}
        \includegraphics[width=0.42\textwidth]{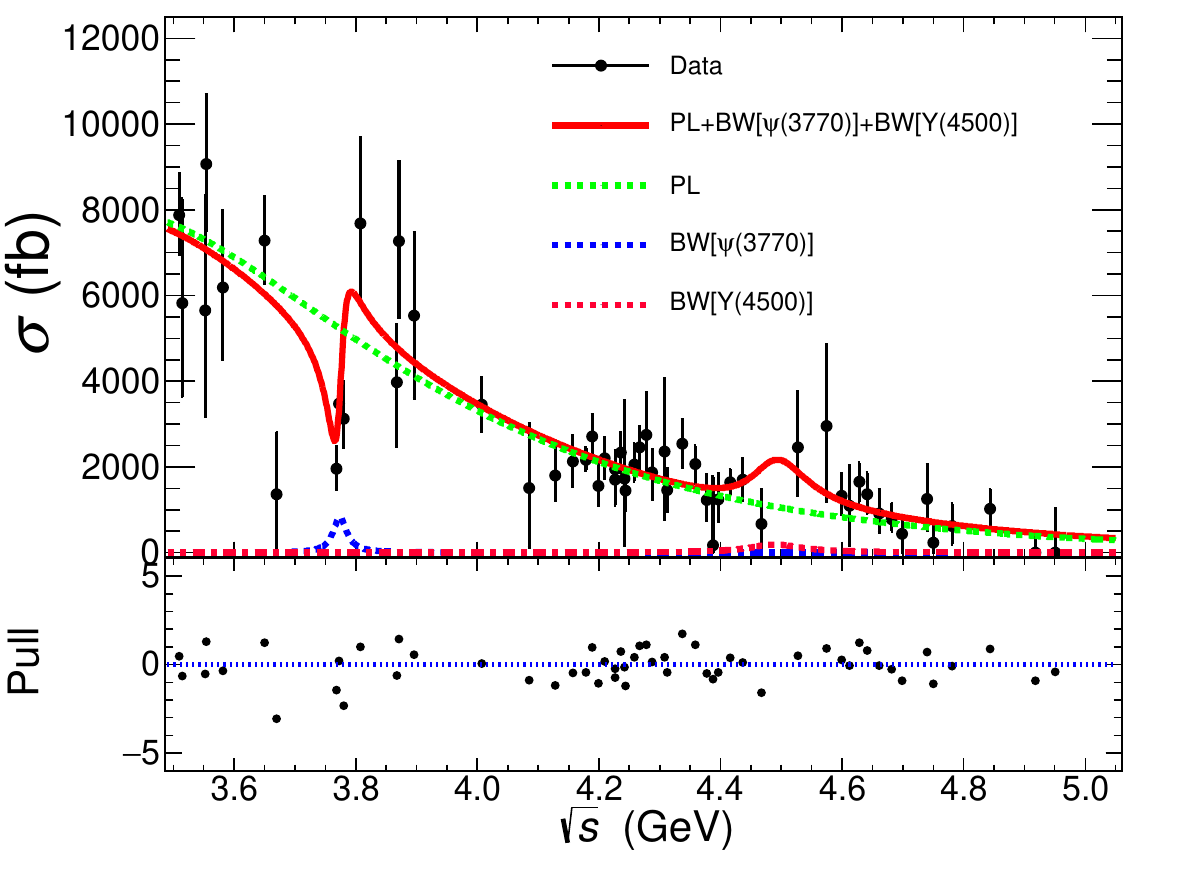}
        \includegraphics[width=0.42\textwidth]{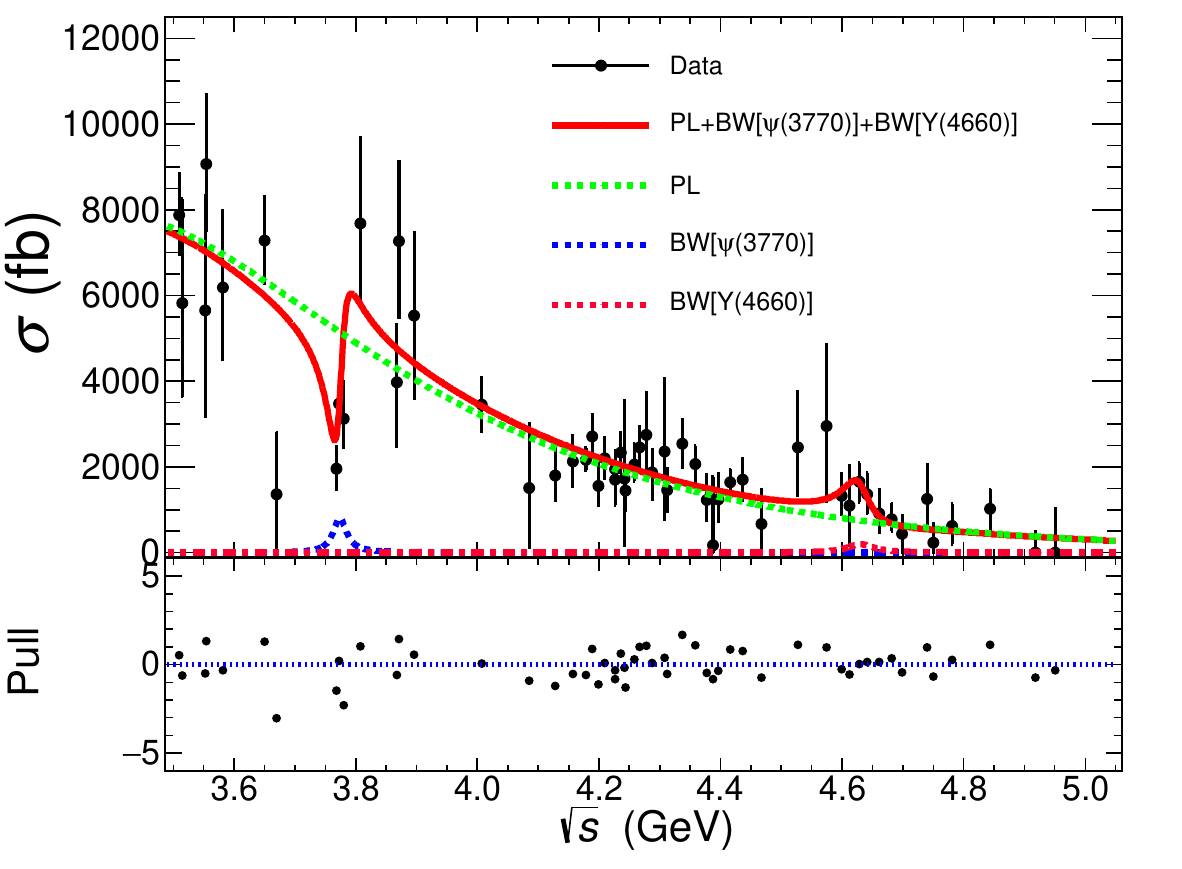}
        \includegraphics[width=0.42\textwidth]{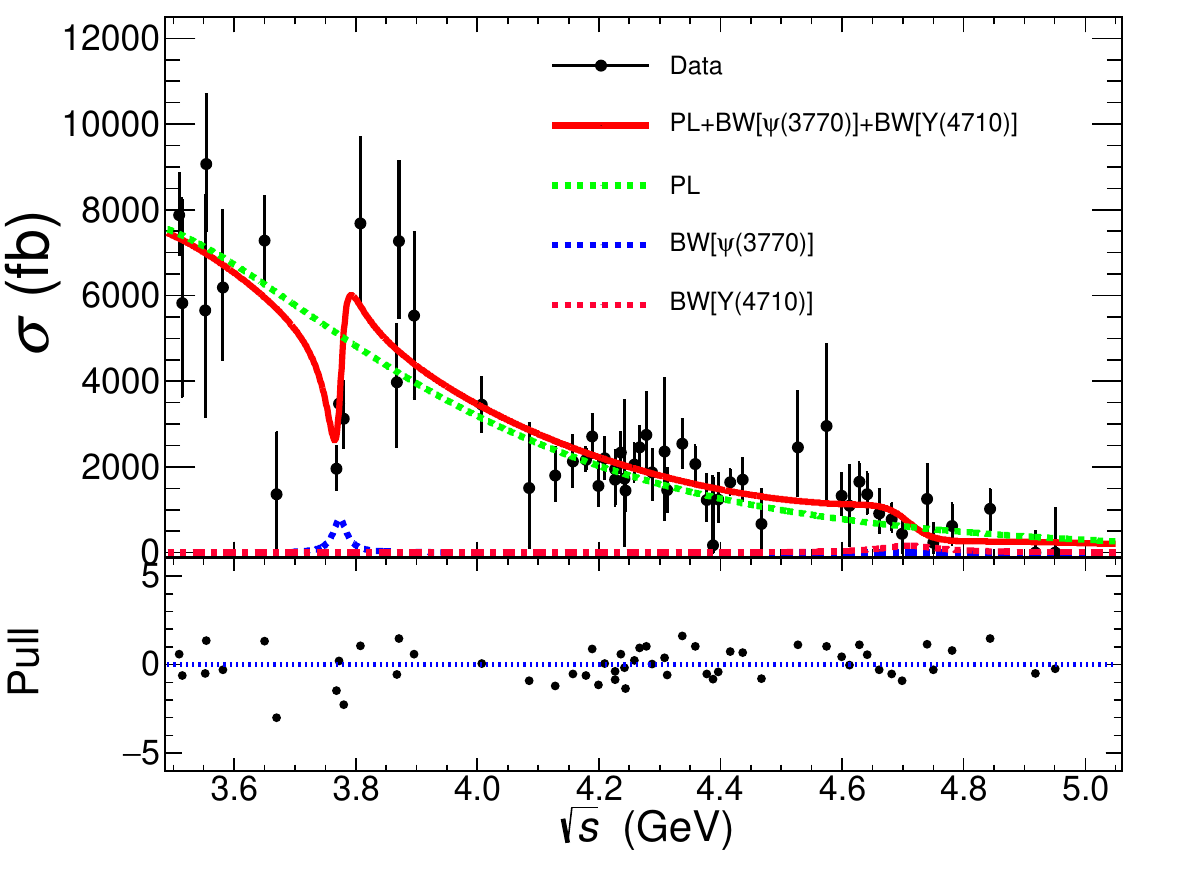}
	\end{center}
    \vspace{-20pt}
	\caption{Fits to the dressed cross sections at c.m.~energies from 3.51 to 4.95~GeV under different resonance assumptions as indicated in the legend. The dots with error bars represent the dressed cross sections and the solid lines indicate the fit results. The error bars consider the statistical and systematic uncertainties summed in quadrature.}
	\label{Fig:XiXi::CS::Line-shape-3773-1}
\end{figure*}

The potential resonances in the cross sections for the $\EE\rightarrow\KXSCC$ reaction are studied by fitting the dressed cross sections, $\sigma^{\rm dressed} =\sigma^{B}/|1-\Pi|^2$, using the least-squares method. The fit minimizes the $\chi^2$ function defined as:
$\chi^2 = \sum_i\frac{(x_i-fk_i)^2}{\sigma_i^2}+\frac{(1-f)^2}{\sigma_f}$, where $f$ and $\sigma_f$ are a scaling factor and relative correlated systematic uncertainty, respectively, $x_i$ and $k_i$ are the measured and expected cross sections at the $i$-th point, and $\sigma_i$ is the uncorrelated systematic uncertainty.
\begin{figure*}[!hbpt]
        \centering
        \includegraphics[width=0.49\textwidth]{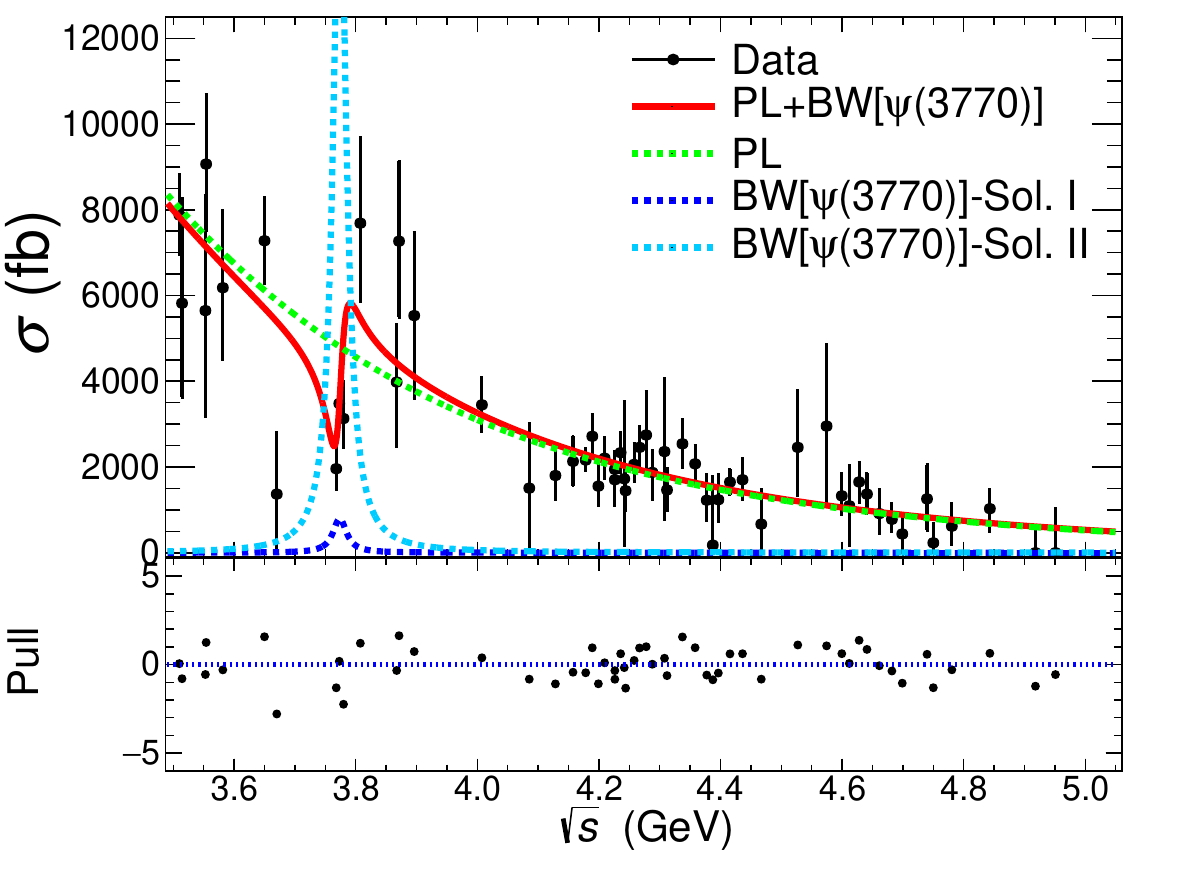}
        \includegraphics[width=0.49\textwidth]{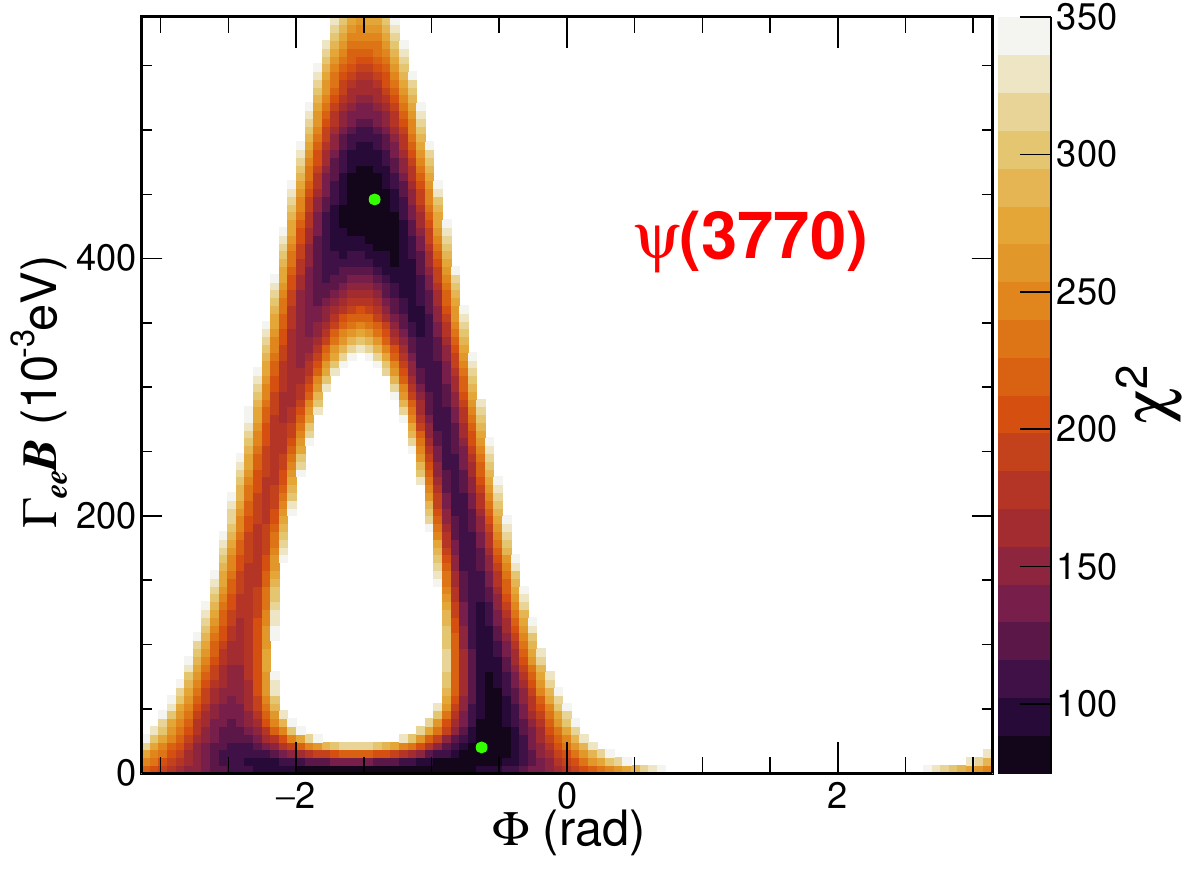}
    \caption{\textbf{Left}: Fit to the dressed cross sections of the $e^+e^-\to\KXSCC$ process. The dots with error bars indicate the measured dressed cross sections including the statistical and systematic uncertainties. The green dashed line corresponds to the fit curve with only the PL function, while the blue dashed line and cyan dashed line represents the two possible solution of signal of the $\psi(3770)$ resonance. Sol. I and Sol. II represent solution I and solution II. The solid red line denotes the fit incorporating both contributions. The bottom panel displays the pull distribution, $(x_i-k_i)/\sigma^{\sigma^B}_i$, with $x_i$ and $k_i$ being the measured and expected cross sections (PL+BW[$\psi(3770)$]), respectively, at the $i$-th point, and $\sigma^{\sigma^B}_i$ being the uncertainty at this point.
    \textbf{Right}: The contour of $\Gamma_{ee}\mathcal{B}$ and $\phi$ on the distribution of $\chi^2$ values. The green points represent the center values from the best fit.}
	\label{Fig:XiXi::CS::Line-shape-3773}
\end{figure*}
The systematic uncertainties from luminosity, kaon tracking and kaon PID, $\Xi^0$ reconstruction, and the input line shape are assumed to be fully correlated among different c.m.~energies, while the other systematic uncertainties are assumed to be uncorrelated. 
Assuming that the cross sections of $\EE\ar\KXSCC$ include a resonance $\psi(3770)$, $\psi(4040)$, $\psi(4160)$, $Y(4230)$, $Y(4360)$, $\psi(4415)$, $Y(4500)$, $Y(4660)$ or $Y(4710)$ and a contribution from the continuum process, a fit to the dressed cross sections is applied with the coherent sum of a power-law~(PL) function and Breit-Wigner~(BW) functions~\cite{Ablikim:2019kkp}:
 \begin{equation}\label{BCS_1}
\begin{split}
\sigma^{\text{dressed}}(\sqrt{s}) 
= \bigg\lvert &\, c_0 \frac{\sqrt{P(\sqrt{s})}}{\sqrt{s}^n} + e^{i\phi} {\rm BW}(\sqrt{s}) \sqrt{\frac{P(\sqrt{s})}{P(M)}} 
\bigg\rvert^2.
\end{split}
\end{equation}
Here, $\phi$ is the relative phase between the BW function
  \begin{equation}
{\rm BW}(\sqrt{s}) =\frac{\sqrt{12\pi\Gamma_{ee}{\cal{B}}\Gamma}}{s-m^{2}+im\Gamma}
 \end{equation}
and the PL function, $c_0$ and $n$ are free parameters, $\sqrt{P(\sqrt{s})}$ is the three-body phase-space factor.
The masses $m$ and the total widths $\Gamma$ for the $Y(4500)$ and the $Y(4700)$ are fixed to the values in Refs.~\cite{BESIII:2022joj, BESIII:2023wqy}, while other possible states are fixed to the PDG values.
The $\Gamma_{ee}{\cal{B}}$ is the product of the electronic partial width and the branching fraction for the possible resonance decaying into the $\KXSCC$ final state. 
Additionally, since the coherent sum of a PL function and BW functions can cause multisolution~\cite{Bai:2019jrb}, the multisolution scan results and the fit results for the $\psi(3770)$ resonance assumption are provided in Fig.~\ref{Fig:XiXi::CS::Line-shape-3773}. The fits to other possible resonances are shown in Fig.~\ref{Fig:XiXi::CS::Line-shape-3773-1}.
The fit results with different resonance hypotheses are summarized in Table~\ref{tab:multisolution}.
Here the $\psi(3770)\to\KXSCC$ decay is observed with a significance of 6.0$\sigma$ after considering the systematic uncertainties.
No significant signals for other possible charmonium(like) states,
\textit{i.e.}~$\psi(4040)$, $\psi(4160)$, $Y(4230)$, $Y(4360)$, $\psi(4415)$, $Y(4500)$, $Y(4660)$ or $Y(4710)$, are found.
The $\Gamma_{ee}{\cal{B}}$ values, ${\cal{B}}$ values and their upper limits including the systematic uncertainty at the 90\% confidence level, computed with a Bayesian approach~\cite{Zhu:2008ca}, are provided. The $\Gamma_{ee}$ values are fixed to the PDG values~\cite{ParticleDataGroup:2022pth}.
The possible multiple solutions for the $\psi(3770)$ resonance parameters in the fit of the dressed cross sections are obtained by scanning $\Gamma_{ee}{\cal{B}}$ and $\phi$ in the parameter space as shown in Fig.~\ref{Fig:XiXi::CS::Line-shape-3773}.

\begin{table*}[!hbpt]
    \caption{The fitted resonance parameters for $\Gamma_{ee}\mathcal{B}~(10^{-3}~\rm{eV})$, $\mathcal{B}~(10^{-6})$ and $\phi$~(rad). The values in first and second columns represent the parameters of $\psi(3770)$, and values in other columns represent the parameters of other resonances. The fit procedure includes both statistical and systematic uncertainties. Sol. I and Sol. II represent two possible solutions of the fit.}
    \centering
    \begin{tabular}{c c c c c}
    \\
        \hline
        \hline
        \small Resonance &$\Gamma_{ee}\mathcal{B}$ $\rm{(10^{-3} eV)}$ &$\mathcal{B}$ $\rm{(10^{-6})}$& $\phi$ $\rm{(rad)}$ &$\chi^2/n.d.f$\\
        \hline
PL&$\cdot\cdot\cdot$&$\cdot\cdot\cdot$
&$\cdot\cdot\cdot$
&100.08/51\\
$\psi(3770)$(Sol. I)&$20.0^{+9.7}_{-7.3}$&$76.5^{+37.4}_{-28.2}$
&$-0.63~\pm~0.08$
&55.95/51\\
$\psi(3770)$(Sol. II)&$446.6^{+21.1}_{-20.9}$&$1704.8^{+142.2}_{-141.6}$
&$-1.42~\pm~0.05$
&55.89/51\\
$\psi(3770)+\psi(4040)$&$75.0^{+87.8}_{-43.7}~(<220.1)$&$87.2^{+102.3}_{-51.4}~(<255.9)$
&$-2.13~\pm~0.16$
&50.42/49\\
$\psi(3770)+\psi(4160)$&$1.7^{+1.9}_{-1.2}~(<6.1)$&$3.6^{+4.3}_{-2.9}~(<12.7)$
&$-1.37~\pm~0.55$
&53.21/49\\
$\psi(3770)+Y(4230)$&$2.4^{+2.3}_{-1.5}~(<8.0)$&$\cdot\cdot\cdot$&$-0.96~\pm~0.33$
&52.29/49\\
$\psi(3770)+Y(4360)$&$4.1^{+4.2}_{-2.7}~(<14.2)$&$\cdot\cdot\cdot$&$\phantom{-}0.95~\pm~0.57$
&53.23/49\\
$\psi(3770)+\psi(4415)$&$4.1^{+4.9}_{-2.9}~(<17.4)$&$11.6^{+14.6}_{-9.3}~(<49.7)$&$\phantom{-}0.79~\pm~0.60$&53.75/49\\
$\psi(3770)+Y(4500)$&$28.8^{+20.8}_{-16.1}~(<77.1)$&$\cdot\cdot\cdot$&$\phantom{-}1.12~\pm~0.24$
&52.25/49\\
$\psi(3770)+Y(4660)$&$15.7^{+10.3}_{-7.4}~(<40.1)$&$\cdot\cdot\cdot$&$\phantom{-}2.22~\pm~0.36$
&49.65/49\\
$\psi(3770)+Y(4710)$&$30.5^{+43.8}_{-20.7}~(<604.5)$&$\cdot\cdot\cdot$&$-3.05~\pm~0.36$
&52.79/49\\
        \hline
        \hline
    \end{tabular}
    \label{tab:multisolution}
\end{table*}

\begin{figure}[h]
\centering
\hspace{-1.5em}
\includegraphics[width=0.48\textwidth]{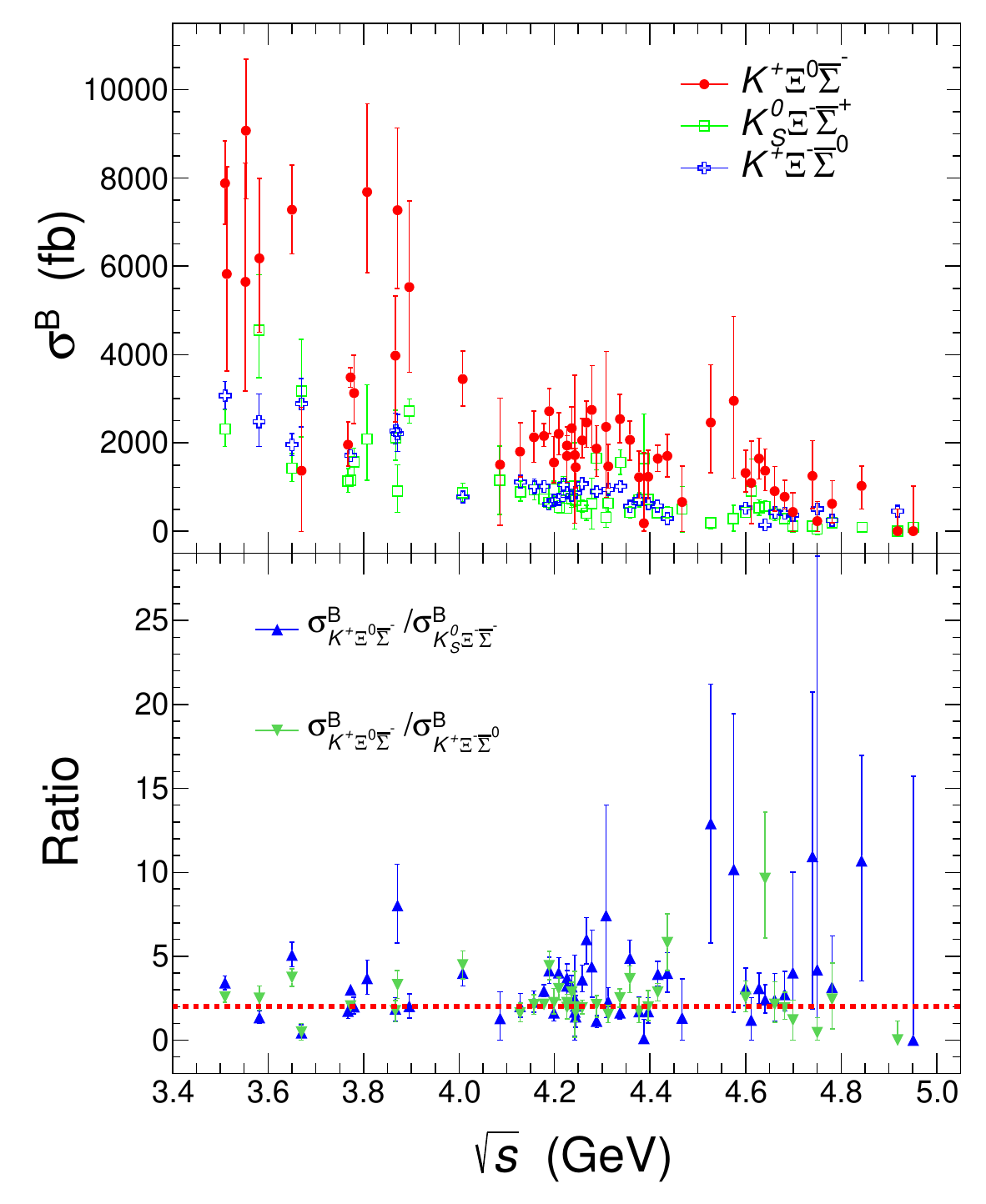}
\caption{The Born cross sections (top) as a function of the c.m.~energy for the $e^+e^-\to\KXSCC$ channel, in comparison to the previous BESIII measurements of $e^+e^-\to \KSXSCC$~\cite{BESIII:2026fcp} and $e^+e^-\to \KXSZCC$~\cite{BESIII:2024ogz}.
The ratios of Born cross sections (bottom) between this work and two isospin partner processes, compared to a rough theoretical prediction of about two (red line), based on isospin conservation.}
\label{Fig:ratio_of_sig}
\end{figure}
In summary, using $\EE$ collision data collected by the BESIII detector at the BEPCII collider corresponding to an integrated luminosity of 44.55~fb$^{-1}$, the Born cross sections for the $e^+e^-\to\KXSCC$ reaction are measured at 56 c.m.~energies ranging from 3.51 to 4.95~GeV.
A fit to the dressed cross sections of the $e^+e^-\to\KXSCC$ reaction is performed under the assumption of a charmonium(like) resonance \textit{i.e.}~$\psi(3770)$, $\psi(4040)$, $\psi(4160)$, $Y(4230)$, $Y(4360)$, $\psi(4415)$, $Y(4500)$, $Y(4660)$ or $Y(4710)$ and a continuum contribution. 
The decay of $\psi(3770)\to\KXSCC$ is observed for the first time with a significance of 6.0$\sigma$, including the systematic uncertainty.
No significant signal for any other charmonium(like) resonances is found. The $\Gamma_{ee}\cal{B}$ values and the upper limits at the 90\% confidence level for these charmonium(like) states decaying into the $\KXSCC$ final state are provided.
In addition, the ratios of Born cross sections of the $e^+e^-\to\KXSCC$ reaction to those of the isospin partner processes $e^+e^-\to \KSXSCC$ and $e^+e^-\to \KXSZCC$ are presented in Fig.~\ref{Fig:ratio_of_sig}. The measured values are in a good agreement with the scenario of isospin conservation.
The obtained results are crucial for studying vector charmonium(like) states decays into charmless baryonic final states, providing important experimental input for investigating both the origin of the large non-$D\bar{D}$ component of the $\psi(3770)$ decay width and the nature of vector charmonium(like) states above the open-charm threshold.

\section{ACKNOWLEDGMENTS}
The BESIII Collaboration thanks the staff of BEPCII (\url{https://cstr.cn/31109.02.BEPC}) and the IHEP computing center for their strong support. This work is supported in part by National Key R\&D Program of China under Contracts No. 2025YFA1613900, No. 2023YFA1606000, and No. 2023YFA1606704; National Natural Science Foundation of China (NSFC) under Contracts No. 12247101, No. 11635010, No. 11935015, No. 11935016, No. 11935018, No. 12025502, No. 12035009, No. 12035013, No. 12061131003, No. 12192260, No. 12192261, No. 12192262, No. 12192263, No. 12192264, No. 12192265, No. 12221005, No. 12225509, No. 12235017, No. 12342502, and No. 12361141819; the Fundamental Research Funds for the Central Universities No. lzujbky-2025-ytA05, No. lzujbky-2025-it06, and No. lzujbky-2024-jdzx06; the Natural Science Foundation of Gansu Province No. 22JR5RA389 and No. 25JRRA799; the “111 Center” under Grant No. B20063; the Chinese Academy of Sciences (CAS) Large-Scale Scientific Facility Program; the Strategic Priority Research Program of Chinese Academy of Sciences under Contract No. XDA0480600; CAS under Contract No. YSBR-101; 100 Talents Program of CAS; The Institute of Nuclear and Particle Physics (INPAC) and Shanghai Key Laboratory for Particle Physics and Cosmology; ERC under Contract No. 758462; German Research Foundation DFG under Contract No. FOR5327; Istituto Nazionale di Fisica Nucleare, Italy; Knut and Alice Wallenberg Foundation under Contracts No. 2021.0174, No. 2021.0299, and No. 2023.0315; Ministry of Development of Turkey under Contract No. DPT2006K-120470; National Research Foundation of Korea under Contract No. NRF-2022R1A2C1092335; National Science and Technology fund of Mongolia; Polish National Science Centre under Contract No. 2024/53/B/ST2/00975; STFC (United Kingdom); Swedish Research Council under Contract No. 2019.04595; and U.S. Department of Energy under Contract No. DE-FG02-05ER41374.

\bibliography{database}


\onecolumngrid
\begin{center}
M.~Ablikim$^{1}$\BESIIIorcid{0000-0002-3935-619X},
M.~N.~Achasov$^{4,c}$\BESIIIorcid{0000-0002-9400-8622},
P.~Adlarson$^{83}$\BESIIIorcid{0000-0001-6280-3851},
X.~C.~Ai$^{89}$\BESIIIorcid{0000-0003-3856-2415},
C.~S.~Akondi$^{31A,31B}$\BESIIIorcid{0000-0001-6303-5217},
R.~Aliberti$^{39}$\BESIIIorcid{0000-0003-3500-4012},
A.~Amoroso$^{82A,82C}$\BESIIIorcid{0000-0002-3095-8610},
Q.~An$^{79,65,\dagger}$,
Y.~H.~An$^{89}$\BESIIIorcid{0009-0008-3419-0849},
Y.~Bai$^{63}$\BESIIIorcid{0000-0001-6593-5665}
% ...
\\
\vspace{0.2cm}
(BESIII Collaboration)\\
\vspace{0.2cm} {\it
$^{1}$ Institute of High Energy Physics, Beijing 100049, People's Republic of China\\
% ...
$^{4}$ Budker Institute of Nuclear Physics SB RAS (BINP), Novosibirsk 630090, Russia\\
% ...
$^{31}$ INFN Sezione di Ferrara, (A)INFN Sezione di Ferrara, I-44122, Ferrara, Italy; (B)University of Ferrara, I-44122, Ferrara, Italy\\
% ...
$^{39}$ Johannes Gutenberg University of Mainz, Johann-Joachim-Becher-Weg 45, D-55099 Mainz, Germany\\
% ...
$^{63}$ Southeast University, Nanjing 211100, People's Republic of China\\
% ...
$^{65}$ State Key Laboratory of Particle Detection and Electronics, Beijing 100049, Hefei 230026, People's Republic of China\\
% ...
$^{79}$ University of Science and Technology of China, Hefei 230026, People's Republic of China\\
% ...
$^{82}$ University of Turin and INFN, (A)University of Turin, I-10125, Turin, Italy; (B)University of Eastern Piedmont, I-15121, Alessandria, Italy; (C)INFN, I-10125, Turin, Italy\\
$^{83}$ Uppsala University, Box 516, SE-75120 Uppsala, Sweden\\
% ...
$^{89}$ Zhengzhou University, Zhengzhou 450001, People's Republic of China\\

\vspace{0.2cm}
$^{\dagger}$ Deceased\\
% ...
$^{c}$ Also at the Novosibirsk State University, Novosibirsk, 630090, Russia\\
% ...

}

\end{center}

\end{document}